\documentclass[
 reprint,
 amsmath,
 amssymb,
 aps,
 prl, %
 superscriptaddress
]{revtex4-2}

\usepackage[utf8]{inputenc}
\usepackage{graphicx}
\usepackage{hyperref}
\usepackage{gensymb}
\usepackage{siunitx}
\usepackage{threeparttable, tablefootnote}
\usepackage{booktabs}
\usepackage{csquotes}
\usepackage[export]{adjustbox}
\usepackage{placeins}
\usepackage{setspace}
\usepackage{enumitem}
\usepackage{verbatim}
\usepackage{mathtools}
\usepackage[thinc]{esdiff}
\usepackage{amsmath}

\usepackage[caption=false]{subfig}
\usepackage[usenames,dvipsnames]{color}
\usepackage{braket}

\usepackage{dcolumn}
\usepackage{bm}

\usepackage[utf8]{inputenc}
\usepackage[T1]{fontenc}
\usepackage{mathptmx}
\usepackage{etoolbox}

\newcommand{\ion}[2]{\ensuremath{^{#2}\mathrm{#1}^+}}

\begin{document}

\preprint{APS/123-QED}

\title{Long-lived metastable-qubit memory}

\newcommand{\MITAffiliation}[0]{Center for Ultracold Atoms, Research Laboratory of Electronics, Massachusetts Institute of Technology, Cambridge, Massachusetts 02139, USA}
\newcommand{\LLAffiliation}[0]{Lincoln Laboratory, Massachusetts Institute of Technology, Lexington, Massachusetts 02421, USA}

\author{Xiaoyang Shi}
\email{shix@mit.edu}
\affiliation{\MITAffiliation}

\author{Jasmine Sinanan-Singh}
\affiliation{\MITAffiliation}

\author{Kyle DeBry}
\affiliation{\MITAffiliation}
\affiliation{\LLAffiliation}

\author{Susanna L. Todaro}
\thanks{Present address:  Oxford Ionics, Oxford, U.K.}
\affiliation{\MITAffiliation}

\author{Isaac L. Chuang}
\affiliation{\MITAffiliation}

\author{John Chiaverini}
\affiliation{\MITAffiliation}
\affiliation{\LLAffiliation}

\begin{abstract}

Coherent storage of quantum information is crucial to many quantum technologies. Long coherence times have been demonstrated in trapped-ion qubits, typically using the hyperfine levels within the ground state of a single ion. However, recent research suggests qubits encoded in metastable states could provide architectural benefits for quantum information processing, such as the possibility of effective dual-species operation in a single-species system and erasure-error conversion for fault-tolerant quantum computing. 
Here we demonstrate long-lived encoding of a quantum state in the metastable states of a trapped ion. By sympathetically cooling with another ion of the same species and constantly monitoring for erasure errors, we demonstrate a coherence time of 136(42) seconds with a qubit encoded in the metastable $5D_{5/2}$ state of a single \ion{Ba}{137} ion. In agreement with a model based on empirical results from dynamical-decoupling-based noise spectroscopy, we find that dephasing of the metastable levels is the dominant source of error once erasure errors are removed.

\end{abstract}
\maketitle


Qubits encoded in metastable states of atomic systems show great potential for quantum information processing~\cite{Toyoda2009, Sherman2013,Schindler_2013,Ma2023,DeBry2023}. Such states provide access to additional electronic manifolds with high isolation from the more generally used ground-state manifold---coherent and dissipative operations may be performed almost completely independently in these different subspaces, leading to novel utilization of trapped ions and neutral atoms for quantum logic.  For instance, the flexibility that comes with access to separate sets of states suggests new architectures for quantum computing that can circumvent some of the challenges presented by dual-species methods~\cite{Allcock2021,Yang2022}, including those stemming from mass differences between the constituent qubit species.

One possible limitation of metastable states for quantum information processing when compared to the often-employed ground states is the set of additional channels for scattering and decay that the former possess. It is therefore of paramount importance to investigate the behavior of metastable-states when storing quantum information, especially in the presence of other atomic systems that are being used for ancillary operations relevant to carrying out quantum algorithms and quantum error correction. Previous investigations have demonstrated few-to-tens-of-millisecond-scale~\cite{Pucher2024, Unnikrishnan2024} to second-scale~\cite{Feng2024} coherence in metastable levels in neutral atoms and ions, respectively, but a validation of metastable-state coherence commensurate with the longer coherence times shown for ground states is a prerequisite for effective use of these multiple manifolds in high-fidelity quantum-state storage and manipulation.

\begin{figure*}[htbp]
    \centering
    \includegraphics[width=0.85\textwidth]{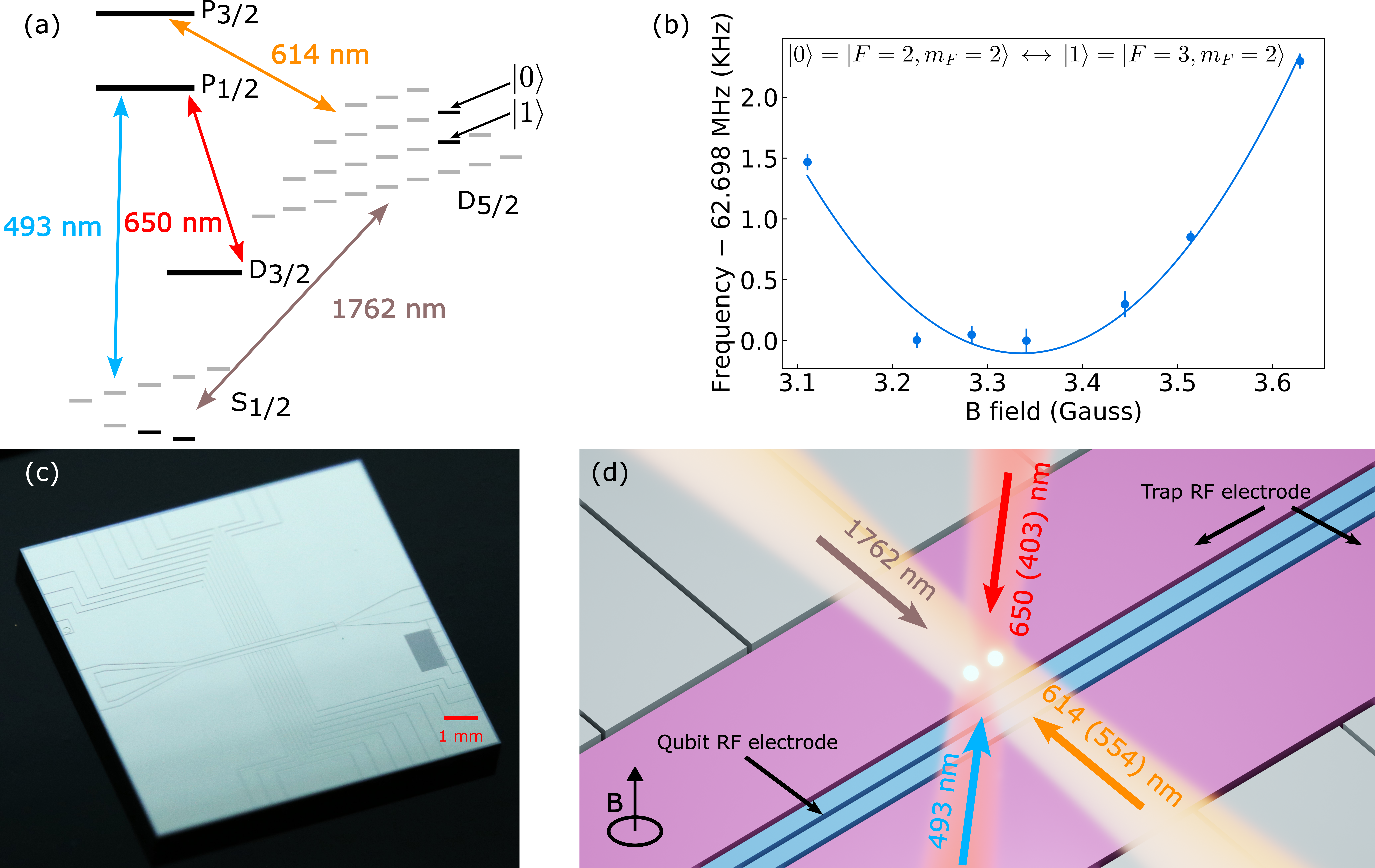}
    \caption{
    Experiment setup and level structure.
    (a) Level diagram of $^{137}$Ba$^+$. The highlighted states in the D$_{5/2}$ and S$_{1/2}$ states are the memory qubit and cabinet shelving states respectively. The hyperfine structure of other levels is neglected for clarity. (b) Metastable qubit transition frequency vs. ambient magnetic field. The solid line is a fit to the data with free parameters being the curvature, an overall frequency shift, and a conversion factor between coil current and magnetic field strength. (c) A photograph of the microfabricated trap used in this work. (d) Surface electrode trap and laser configuration. The electrodes for applying RF currents to drive transitions between hyperfine states are marked in blue and the trapping RF electrodes are marked in purple. The DC electrodes are marked in grey. The laser propagation directions are labeled by arrows. Lasers propagating in the same direction are combined and labeled with parentheses. All beams address both ions equally. }
    \label{fig:combined_setup}
\end{figure*}

Interestingly, the same aspects of metastable states that may lead to the above-mentioned limitations can also provide new enhancements not straightforwardly present in ground states.  The finite state lifetimes of metastable levels lead to decay from the qubit space---however, since this decay is typically to the ground-state manifold, which can be simultaneously used for quantum state detection with isolation from the metastable levels, a decay event can be detected.  Moreover, this leakage event from the metastable qubit space can be converted to an erasure error~\cite{Kang2023, Wu2022, Ma2023}, an error in a specific location, since the specific ion or atom that decayed will be identifiable, and could subsequently be returned to the qubit subspace~\cite{Yang2022,Kang2023}.  While decay of qubit states is in general undesirable, in light of the potential benefits of stimulated-Raman-based quantum logic~\cite{Moore_2023}, using metastable qubits in tandem with erasure conversion of Raman scattering errors can nonetheless be advantageous. Specifically, this capability strengthens the prospect of quantum fault tolerance, as erasure quantum error correcting codes can have very high thresholds, due to the fact that erasure errors are easier to detect than more general Pauli errors~\cite{Grassl1997}.  Demonstrating that such erasure-error detection is possible while maintaining coherence is therefore a key development toward realizing the promise of metastable-qubit processing. 

Here we demonstrate the storage of a quantum state in the metastable D$_{5/2}$ level of \ion{Ba}{137} while sympathetically cooling the memory ion using the ground-state manifold of another ion of the same species.  We also simultaneously monitor the fluorescence from the memory ion to check for leakage errors.  Since neither the sympathetic cooling light nor the light used for resonance-fluorescence-based state detection couples strongly to the memory-qubit manifold, this error detection technique can be effected in real-time to identify decay events.  By proceeding with memory qubit analysis conditioned on the detection of no decay event, a coherence time of 136(42) seconds is observed, approximately four times the metastable state's lifetime, and comparable to ground-state memory times.  The use of mid-algorithm measurement allows us to observe leakage errors explicitly to extrapolate the expected qubit lifetime that an erasure error correction scheme should provide, and hence validates the use of metastable states for high-fidelity quantum logic more generally.

While we use post-selection to evaluate only experimental trials in which the metastable qubit does not decay to the ground state, we point out that the leakage error detection method demonstrated here is generally useful, well beyond post-selected experiments.  Crucially, since the process of monitoring for leakage errors does not affect the memory qubit, the signal from the erasure detection can be used in real-time to verify that no error has occurred~\cite{vizvary2024} and to continue a quantum algorithm; this is to be contrasted with post-selection in a system in which measurement outcomes can be acted upon only after the quantum information is destroyed.  A key consequence is that the long coherence time measured here could be realistically utilized for quantum information processing if leveraged as part of an erasure-based error-correcting code.

As we utilize OMG~\cite{Allcock2021}/dual-type~\cite{Yang2022} encoding in these experiments, individual addressing is required to change the qubit type of specific ions starting in the nominally same state after ion initialization.  This is achieved here using a simple micromotion-assisted individual addressing technique and projection.  Furthermore, metastable-qubit transition energies are subject to environmental perturbations, chief among these being magnetic field fluctuations.  We mitigate dephasing from this source by encoding in a field-insensitive qubit, by surrounding the ion with a superconducting shield, and by applying dynamical decoupling to the memory qubit.

\begin{figure*}[bt]
    \centering
    \subfloat[]{
     \includegraphics[width=0.85\textwidth]{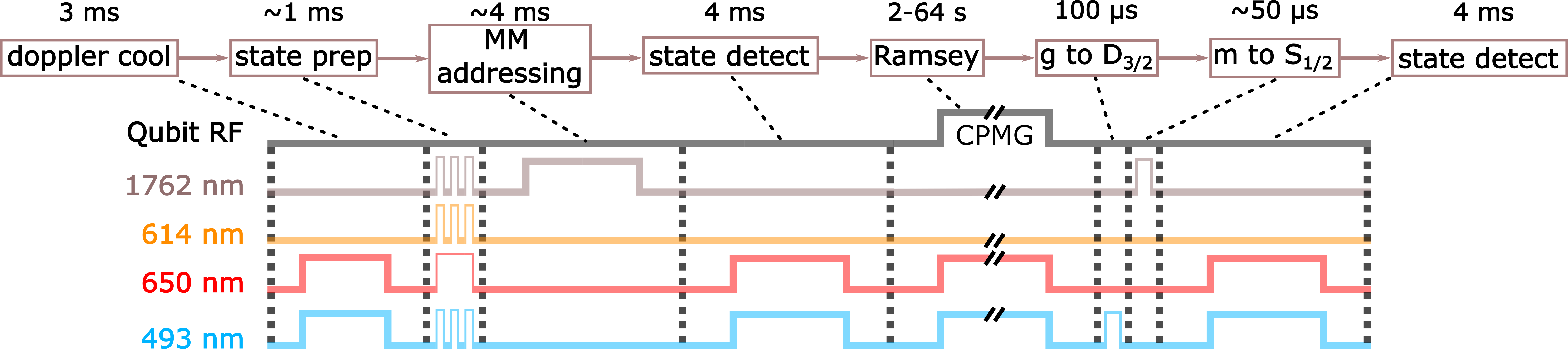}
     \label{fig:pulse_sequence}}\qquad   
     \subfloat[]{
     \includegraphics[width=0.4\textwidth]{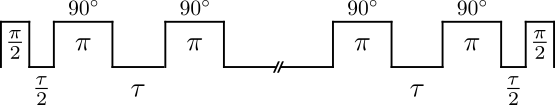}
     \label{fig:pulse_sequence_b}}  
     \caption{Experiment sequence and dynamical decoupling scheme. (a) Eight-step operation sequence for sympathetically cooled quantum memory experiment. "MM": micromotion; "g": ground state qubit; "m": metastable qubit. (b) The CPMG sequence. $\pi$ pulses with 90$^{\circ}$ phases relative to the initial $\pi/2$ pulse are applied interspersed with a delay time $\tau$ which sets the filter function that suppresses (or samples) the noise spectrum.}
    
\end{figure*}

{\bf Experiment}. We use $^{137}$Ba$^{+}$ ions for both the memory and the ancilla ion, with the relevant level structure shown in Fig.~\ref{fig:combined_setup}(a). The memory qubit is encoded in the metastable D$_{5/2}$ level, which has a lifetime of $T_{D}=$30.14(30) s \cite{Zhang2020}. Hyperfine sublevels $\ket{F = 2, m_F = 2}$ and $\ket{F = 3, m_F = 2}$ are the $\ket{0}$ and $\ket{1}$ qubit states, respectively. Based on the measured properties of this species~\cite{Lewty2013}, we expect the qubit transition frequency to be approximately 62.7~MHz and first-order insensitive to magnetic field fluctuation at a magnetic field of 3.34 G. To verify the magnetic field sensitivity, we perform spectroscopy at different ambient fields and the result is shown in Fig.~\ref{fig:combined_setup}(b). The second-order field sensitivity is calculated to be 385 Hz/Gauss$^2$.

As shown in Fig.~\ref{fig:combined_setup}(c,d), we use a microfabricated, surface-electrode, RF Paul trap in a cryogenic system (see more detail in \cite{Sage2012}) for confining the ions. The ion-electrode distance is approximately 50 um and the trap RF drive is at 36 MHz. The trap is cooled to 6~K. For driving RF transitions within the D$_{5/2}$ manifold, we can apply an RF current of a few milliamperes through the middle electrodes (marked in blue), with the other end of the electrodes shorted to ground. Only one of these electrodes is used in this manner for the results presented here. 
Two coils mounted on the intermediate (50~K) stage of the cryocooler in a near-Helmholtz configuration are used for generating a magnetic field perpendicular to the trap surface. There are four rings of niobium (superconducting critical temperature approximately 9.2~K) shielding placed co-axially to the trap-chip normal above and below the chip, with central circular holes of various sizes. These rings significantly suppress magnetic field variation along the vertical direction (chip-normal). More information about, and characterization of, the niobium shielding can be found in the supplementary material, Sec.~\ref{app:shield}.

For loading ions, two photons at 554~nm and 403~nm \cite{Greenberg2023} ionize atoms produced by an effusive oven. For Doppler cooling and state detection, the 493~nm laser is frequency-modulated using an electric-optical modulator (EOM). The laser frequency is set so that the carrier drives the S$_{1/2} \ket{F = 2}$ to P$_{1/2} \ket{F = 2}$ transition, and two microwave tones at approximately 1.47 GHz and 4.0 GHz drive the other three transitions between the hyperfine-split S$_{1/2}$ and P$_{1/2}$ levels. The two tones are switchable using RF switches to allow for state preparation~\cite{An2022} in a particular sublevel. For repumping population from the D$_{3/2}$ state, the 650-nm laser is also modulated by an EOM with two tones so that, along with the carrier, these tones drive the F = $2\xleftrightarrow{}1, 1\xleftrightarrow{}1, 0\xleftrightarrow{}1, 3\xleftrightarrow{}2$ transitions between the D$_{3/2}$ and P$_{1/2}$ level.

For the 614~nm laser, which is used to quench the D$_{5/2}$ state, two acousto-optic modulators (AOMs) set up in a double-pass configuration are used for frequency modulation and switching. We find that it is sufficient to have a single frequency tone for effective quenching of all the sublevels in the D$_{5/2}$ state. An optical shutter is added to the beam path for higher extinction of the 614~nm laser beam during memory-qubit coherence measurements, to avoid unwanted AC Stark shifts and quenching of the D$_{5/2}$ state. A 1762~nm laser locked to a high finesse cavity (StableLaser Systems) and amplified by a thulium-doped fiber amplifier (Cybel, Perseus) is used for state preparation and pre-readout shelving of the memory qubit.

 For experiments with two ions, the pulse sequence is shown in Fig.~\ref{fig:pulse_sequence}. The ions are first Doppler-cooled using the 493 nm and the 650 nm laser beams. We then prepare both ions in the S$_{1/2}\ket{F = 1, m_F = 0}$ state using a similar scheme as in \cite{An2022}, except we do not perform a polarization-based step with the 493 nm laser. To have a consistent assignment of cooling and memory ion, a micromotion-assisted individual addressing technique is used (see supplementary material, Sec.~\ref{app:mm}) to excite the memory ion to the D$_{5/2}$ state. State detection is conducted to verify that only one ion is shelved to the D$_{5/2}$ state. The process is repeated until a one-ion-dark event is detected.

To measure the memory-qubit coherence time, an RF current is sent to the through-wire electrode to drive the transition between the $\ket{0}$ and $\ket{1}$ states. The radio-frequency signal is generated via direct digital synthesis (DDS, M-labs, Urukul) synchronized to an ovenized crystal oscillator, and we drive pulses to implement the CPMG dynamical decoupling sequence~\cite{meiboom1958modified}, a modified Ramsey sequence (see Fig.~\ref{fig:pulse_sequence_b}).  The Doppler cooling beams are kept on during the Ramsey sequence for sympathetic cooling and erasure error detection. To detect leakage errors, we make a measurement of scattered photons using a photomultiplier tube every 4.05 ms to check if the memory ion decayed from the D$_{5/2}$ state. The sequence is terminated when spontaneous decay is detected via fluorescence from both ions.

After the Ramsey sequence, the qubit $\ket{0}$ state is shelved back to the ground state for state detection. To avoid shelving the cooling ion to the D$_{5/2}$ state during the same operation, the 493~nm laser is first switched on for 1~ms to shelve the cooling-ion population into the D$_{3/2}$ state. Two ``cabinet shelving'' pulses~\cite{An2022} are then used to drive the $\ket{0}$ population to the S$_{1/2} \ket{F = 1, m_F = 0, 1}$ states to achieve a higher shelving fidelity. The state of the memory qubit is then extracted by fluorescence detection (the cooling ion remaining dark).

\begin{figure}[!t]
\includegraphics[width=0.47\textwidth]{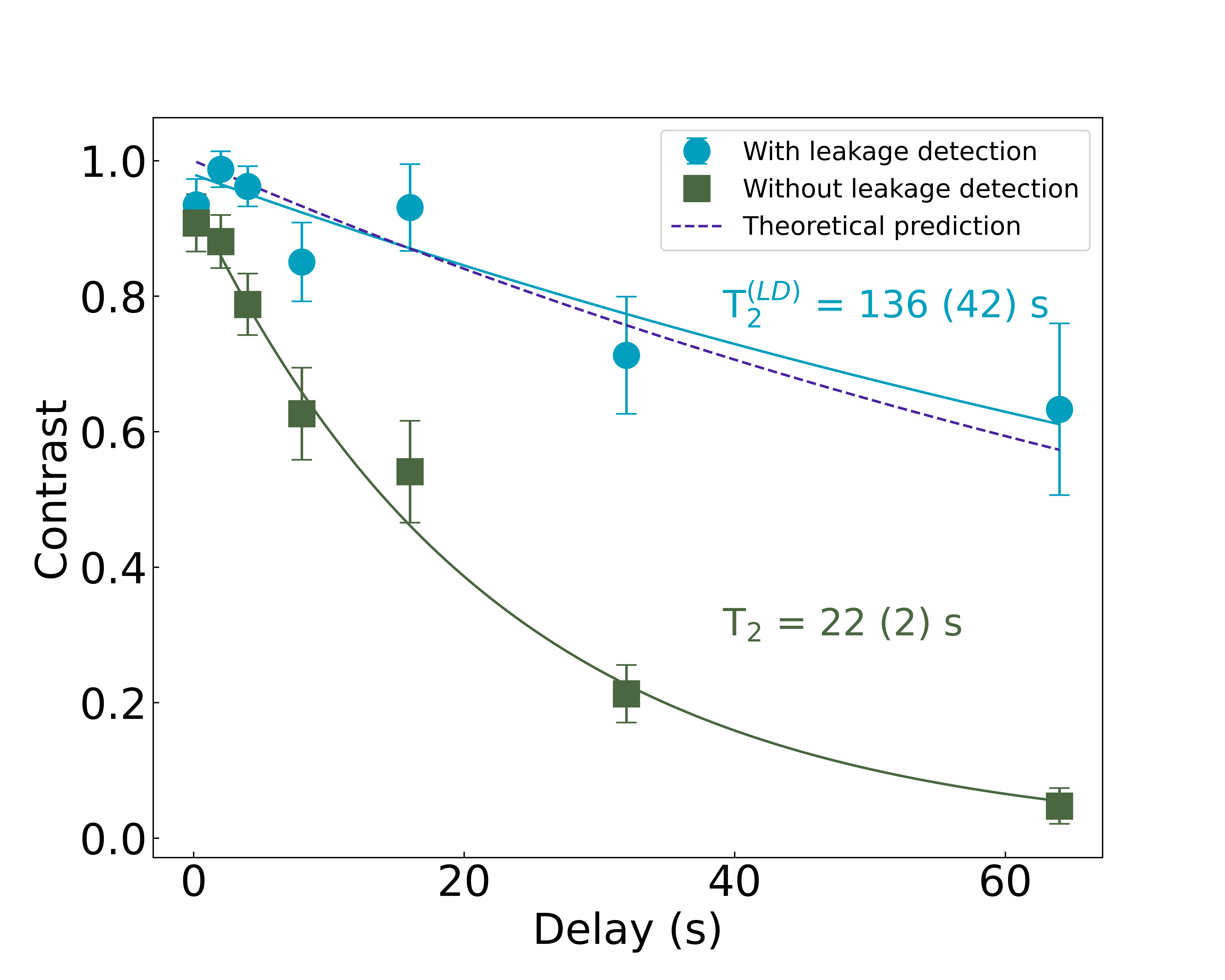}
\caption{ 
Experimental measurement of qubit coherence with no erasure detection (green squares), with real-time erasure detection (blue circles), and expected coherence time based on measured noise spectrum (purple dashed line, see Sec.~\ref{app:noise-modelling}). A coherence time of 136(42) seconds is measured with erasure error detection, while 22(2) seconds is measured without erasure error detection, limited by decay from the qubit states (see text). Error bars correspond to one standard deviation.  Solid lines are exponential decay fits to the data points.}
\label{fig:ramsey_135s}
\end{figure}

{\bf Results.}  The result of the coherence time measurement with leakage detection $T_{2}^{(LD)}$ is shown in Fig.~\ref{fig:ramsey_135s}. A coherence time of 136(42)~s is extracted from an exponential fit. We also perform the experiment without leakage detection, and the coherence time in that case, $T_{2}$, is measured to be 22(2)~s.  These two coherence times taken together are consistent with the lifetime of the D$_{5/2}$ state $T_{D}$ if we assume that the coherence time of the metastable qubit in the presence of decay ($T_{2}$) is dependent on the combination of decay from the qubit states and pure dephasing, the latter characterized by a time $T_{\phi}$.  In that case, we would assume the relationship $1/T_{2}\geq 1/T_{1}+1/T_{\phi}$ (note that the appropriate decoherence rate due to state decay here is $1/T_{1}$ rather than $1/(2T_{1})$ since both qubit states can decay).  For $T_{\phi}=T_{2}^{(LD)}=136(42)$~s and $T_{1}=T_{D}$, we obtain $T_{2}\leq 24.6(7.6)$~s for the decoherence time in the presence of leakage, consistent with the measured 22(2)~s.    

To understand the limits to the measured coherence time with leakage detection, we model the dephasing as follows: fluctuations of the qubit energy splitting are captured by including a dephasing term $\beta(t)$ in addition to the bare qubit splitting $\omega_{0}$ in the Hamiltonian $H = \frac{\hbar}{2} \big(\omega_{0} + \beta(t)\big) \sigma_z$, where $\sigma_z$ is the usual Pauli matrix.  We employ a set of CPMG pulse sequences, each consisting of a series of $N$ total $\pi$ pulses spaced by time $\tau$ to empirically estimate the frequency noise spectrum $S(\omega)$ associated with the noise process $\beta(t)$ via qubit-noise spectroscopy\cite{Wang2017}.

\begin{figure}[!t]
\includegraphics[width=0.99\columnwidth]{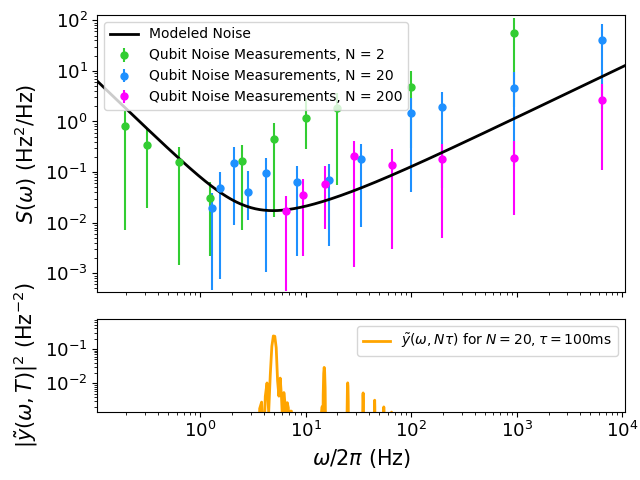}
\caption{Qubit-noise spectroscopy and fitted qubit-splitting frequency-noise spectrum.  The green, blue, and pink data points signify noise amplitudes based on coherence measurements of the metastable qubit using CPMG sequences of varying length and number of pulses $N$.  The noise spectrum, modeled as described in the text, is a fit to these data.  The effective filter function of the CPMG sequence used in the coherence-time measurement shown in Fig.~\ref{fig:ramsey_135s} is depicted in the lower panel.}

\label{fig:modelled-noise}
\end{figure}

The effect of applying a CPMG sequence is similar to applying a filter to the noise spectrum with a center frequency of $\omega_c = 2\pi \times\frac{1}{2\tau}$ and width $1/T$, for total delay time $T=N\tau$, as described by a frequency domain filter function $|\Tilde{y}(\omega, T)|$.  We perform experiments for a set of CPMG sequences with varying $N$ and measure the coherence decay $\chi(T)$:
\begin{equation}
    \chi(T) = \int_{0}^{\infty}S(\omega)|\Tilde{y}(\omega, T)|^2 d\omega
    \label{eq:chi-cpmg}
\end{equation}
\noindent (see Sec.~\ref{app:cpmg} for a full derivation).  We fit \eqref{eq:chi-cpmg} with experimental coherence decay data points $\{ \chi_{\rm exp}^{(N)}(T) \}$ to extract $S(\omega)$ at measured total delays as shown in Fig.~\ref{fig:modelled-noise}.  We fit a power-law spectrum to the data (see Sec.~\ref{app:noise-modelling} for fitting details) and find that the noise spectrum has a minimum at $\omega\approx2\pi \times 5$ Hz and increases rapidly at both lower and higher frequencies. Using this spectrum and the applied CPMG sequence parameters for the coherence time measurement ($\tau=0.1$~s), we predict a coherence decay as shown by the dashed line in Fig.~\ref{fig:ramsey_135s}, consistent with the measured coherence.

{\bf Discussion.}  Environmental noise and imperfections in control fields both contribute to the observed noise spectrum. Potential high-frequency noise may arise from phase noise in the qubit drive, performed via direct digital synthesis of an RF current delivered to the excitation electrodes, though separately measured phase noise is too low to account for the full noise amplitude.  Another possible source of high-frequency noise is magnetic-field fluctuations: the attenuation provided by the superconducting shield decreases with increasing frequency \cite{Kamitani2000}.

One major source of low-frequency noise is fluctuation in the amplitude of the applied RF voltage used for ion confinement.  The asymmetry of the trap-electrode structure in the vertical ($z$) direction leads to uncancelled magnetic fields arising from the current in the trap-RF electrodes. Due to the small frequency difference between the trap RF frequency and the qubit frequency, this magnetic field off-resonantly couples to the qubit and causes AC Zeeman shifts. Fluctuations of the RF amplitude thus lead to qubit-frequency fluctuations. Variation of the trapping field also induces frequency fluctuation due to the quadrupole moment of the D$_{5/2}$ state~\cite{Itano2000}, but as shown in the supplementary material (see Sec.~\ref{app:RF_fluc}), this effect does not contribute significantly in our case. Noise caused by the trap-RF voltage could be suppressed by the addition of active stabilization of its amplitude~\cite{Johnson2016}. Additional low-frequency noise may be caused by magnetic-field fluctuations in spatial directions with insufficient shielding; more complete shielding could potentially reduce such noise at the cost of reduced optical access.

Decoherence also arises from spontaneous Raman photon scattering from the Doppler cooling light~\cite{Ozeri_2005}, which is also applied during the memory storage period, but this effect is negligible due to the large detunings from relevant transitions (see Sec.~\ref{app:scattering-rate}).

The coherence time observed here for clock states in metastable levels of a trapped ion is comparable in scale to those achievable in the ground-state manifolds of atoms.  This suggests that such a state encoding could be useful for quantum computing in OMG-style architectures~\cite{Allcock2021}, for instance, using multiple $^{137}$Ba$^{+}$ ions as a register.  

As shown here, select ions could be addressed to encode metastable-state qubits, while other ions left in the ground-state manifold could be used for sympathetic cooling, allowing simultaneous periodic erasure-error detection and qubit readout.  Such a mid-circuit measurement capability, as demonstrated here, means quantum error correction could also be applied, providing significant functionality in a single-species system.  This demonstration of several key OMG-architecture elements thus provides concrete steps toward the realization of trapped-ion quantum computing at scale.

\section*{Acknowledgements}

I.\,L.\,C. acknowledges support by the NSF Center for Ultracold Atoms.  This research was supported by the U.S. Army Research Office through grant W911NF-20-1-0037.  Support is also acknowledged from the U.S. Department of Energy, Office of Science, National Quantum Information Science Research Centers, Quantum Systems Accelerator, under Air Force Contract No. FA8702-15-D-0001. Any opinions, findings, conclusions or recommendations expressed in this material are those of the author(s) and do not necessarily reflect the views of the Department of Energy.

\bibliography{references}

\appendix

\section{Supplementary material}

\begin{figure*}[bt]
    \centering
    \subfloat[]{
    \includegraphics[width=0.42\textwidth]{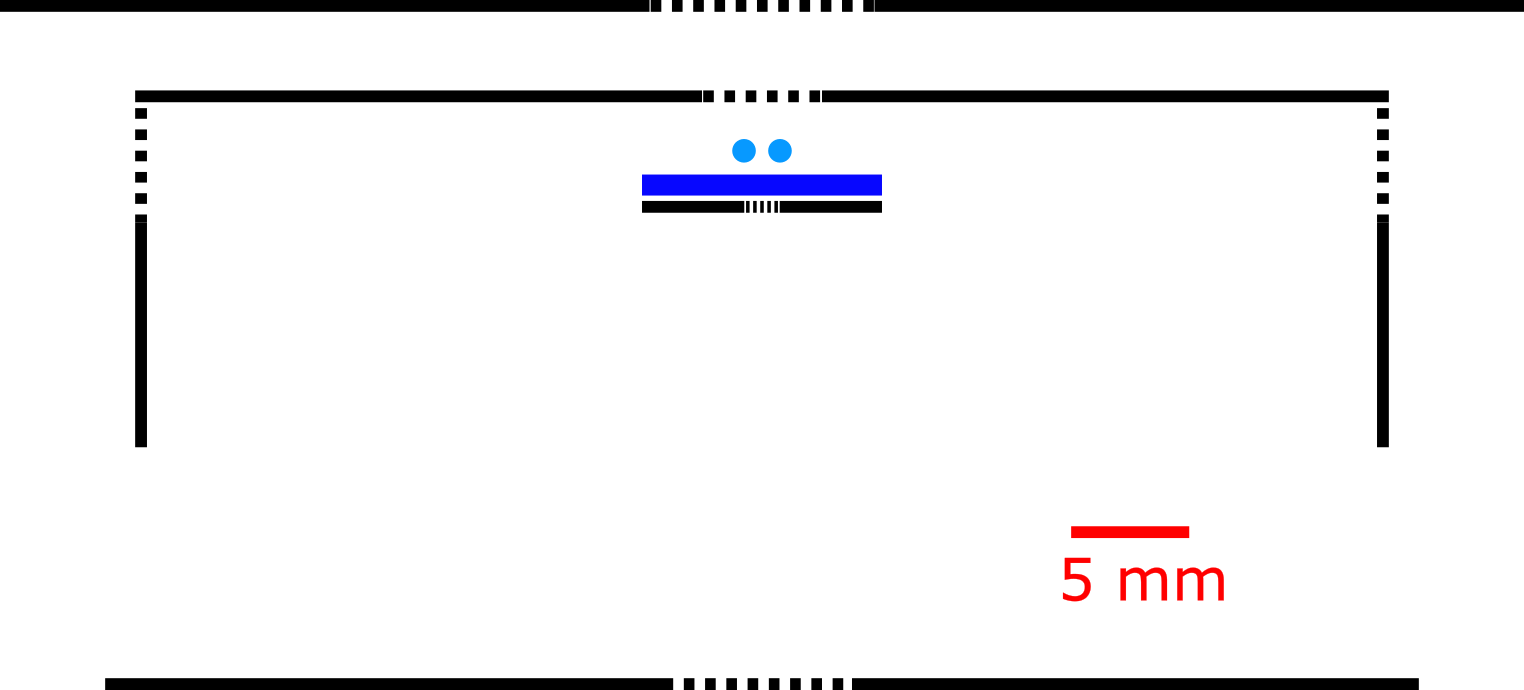}}\qquad   
    \subfloat[]{
    \includegraphics[width=0.35\textwidth]{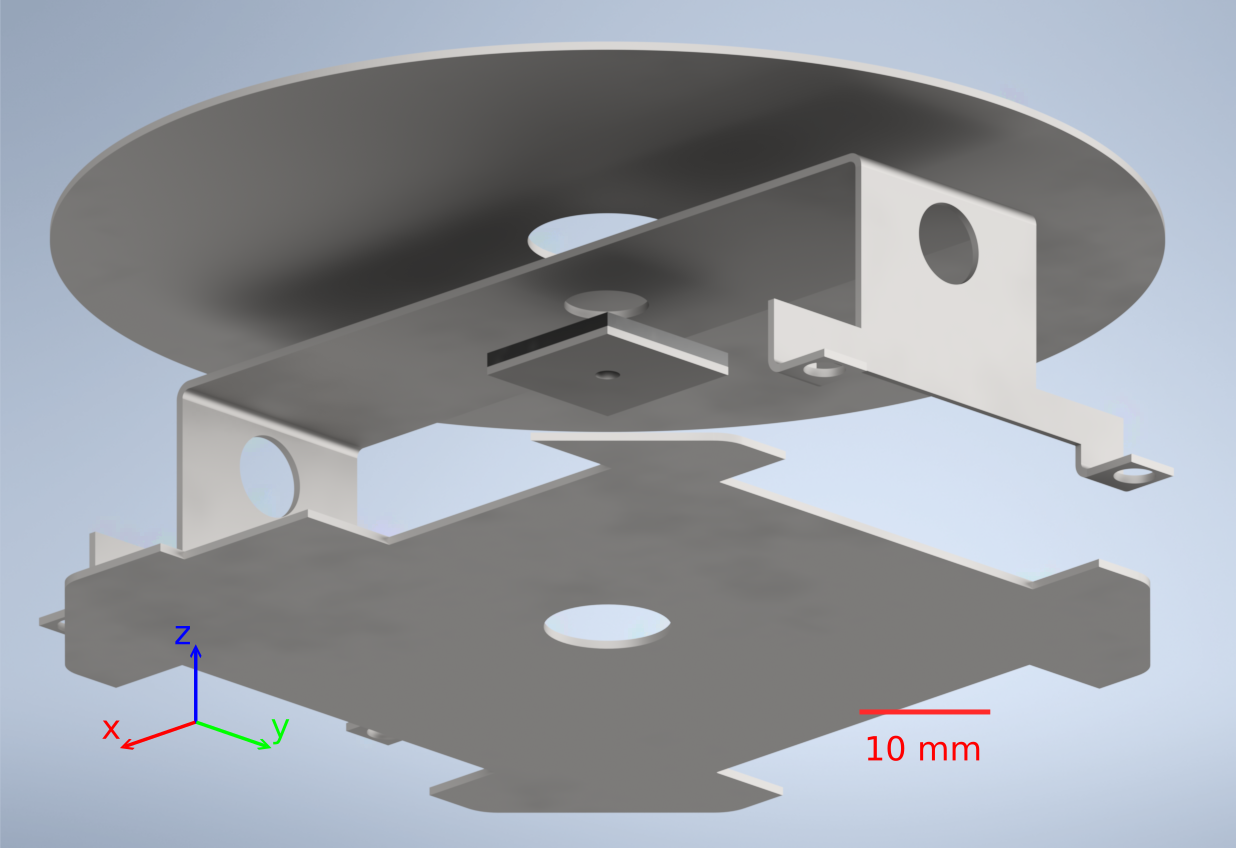}}  
    \caption{The niobium shielding configuration. (a) Schematics of the relative positions of the niobium shields. Black solid lines represent the niobium plates and the dashed line represents the position of a hole. The blue rectangle marks the trap chip and the cyan circles represent the ions (not to scale). (b) In this CAD rendering of the niobium shielding, the trap is the black chip. The 6 K thermal shield and trap mount are not shown.}
    \label{nb_configuration}
\end{figure*}

\subsection{Superconducting shields for reduction of magnetic-field fluctuations}
\label{app:shield}

\begin{figure}[!ht]
    \includegraphics[width=3.0in]{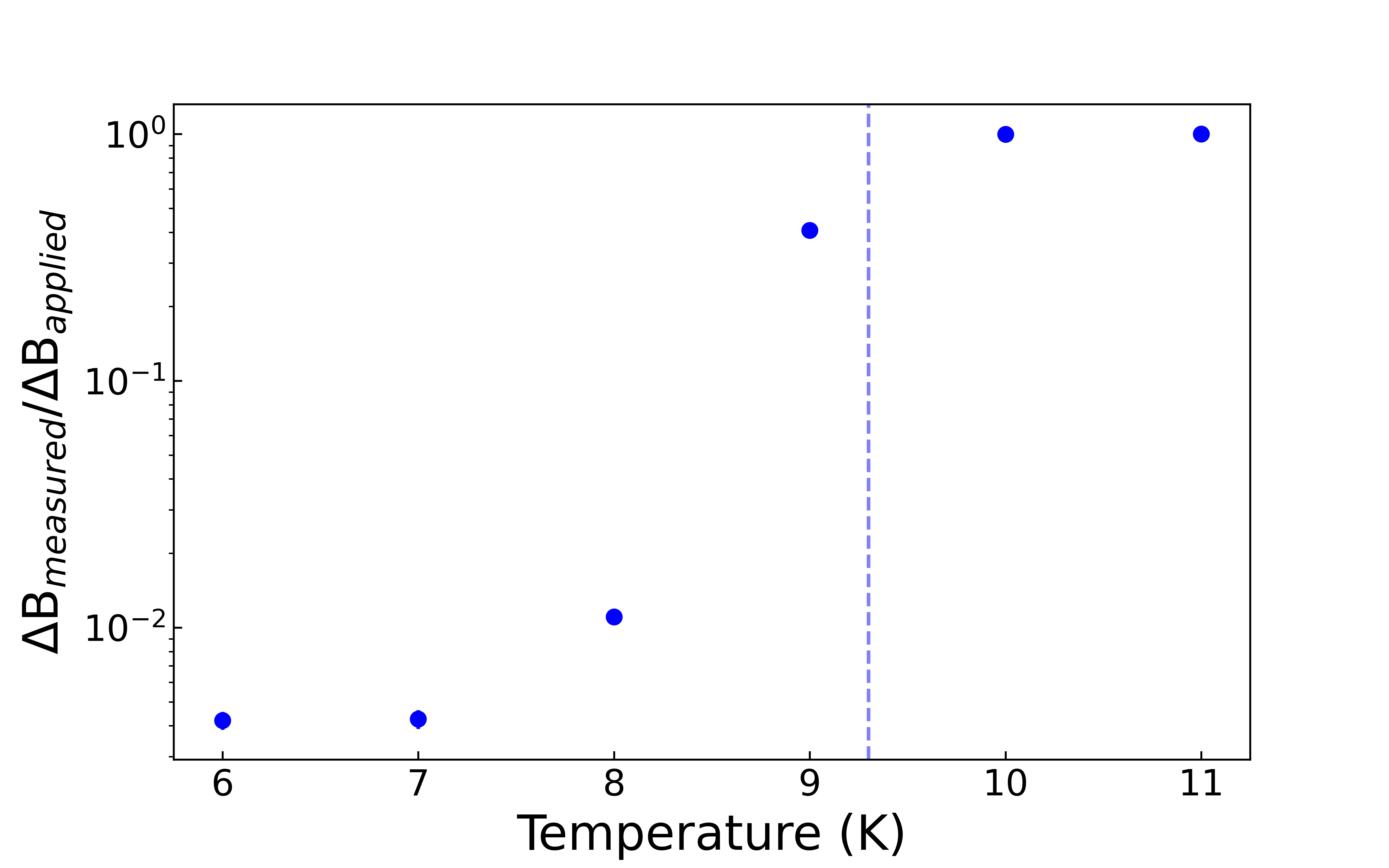}
    \caption{Ratio of measured magnetic field change to applied magnetic field change at different temperatures. The externally applied magnetic field is varied by $\sim 1.4$ Gauss and the local change in the magnetic field is measured via ion spectroscopy as a function of trap-stage temperature. The dashed blue line marks the critical temperature of niobium (approximately 9.2 K). The ambient field change is suppressed by a factor of $\sim$240 at 6~K. Error bars are smaller than the markers.}
    \label{fig:suppresion_factor}
\end{figure}

The niobium shield configuration is shown in Fig.~\ref{nb_configuration}. Four pieces of niobium sheet with various shapes are mounted around the ions to suppress magnetic field variation. To measure the amount of suppression, we choose two transitions ($\ket{F = 1, m_F = 0}$ to $\ket{F = 1, m_F = \pm1}$) between the S$_{1/2}$ and D$_{5/2}$ states for spectroscopy. These two transitions are similarly sensitive to magnetic field changes, but with opposite sign. The difference in the two transition frequencies thus provides a measure of the magnetic field at the ion position. We then perform spectroscopy at two different ambient fields by changing the current applied to the magentic-field coils and measuring the field change ($\Delta$B) as measured at the ion location. The difference between the two applied magnetic fields is approximately 1.4 G. This measurement is repeated at different temperatures of the trap stage, and the result is shown in Fig.~\ref{fig:suppresion_factor}. A suppression factor of ${\sim}$240 is measured for the temperature used in the experiments presented in the main paper (6~K).

\subsection{Micromotion-assisted individual addressing}
\label{app:mm}

\begin{figure*}[!ht]
    \centering
    \subfloat[]{\raisebox{14pt}{
    \includegraphics[height=1.6in]{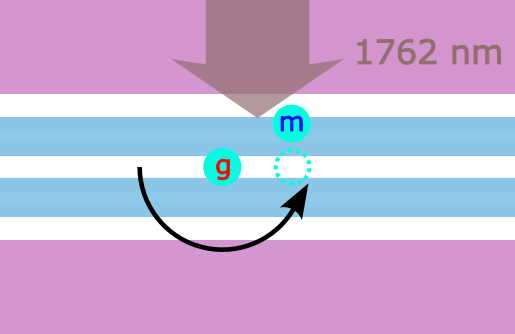}}
    \label{fig:twisting_chain}}\qquad 
    \subfloat[]{
   \includegraphics[height=1.8in]{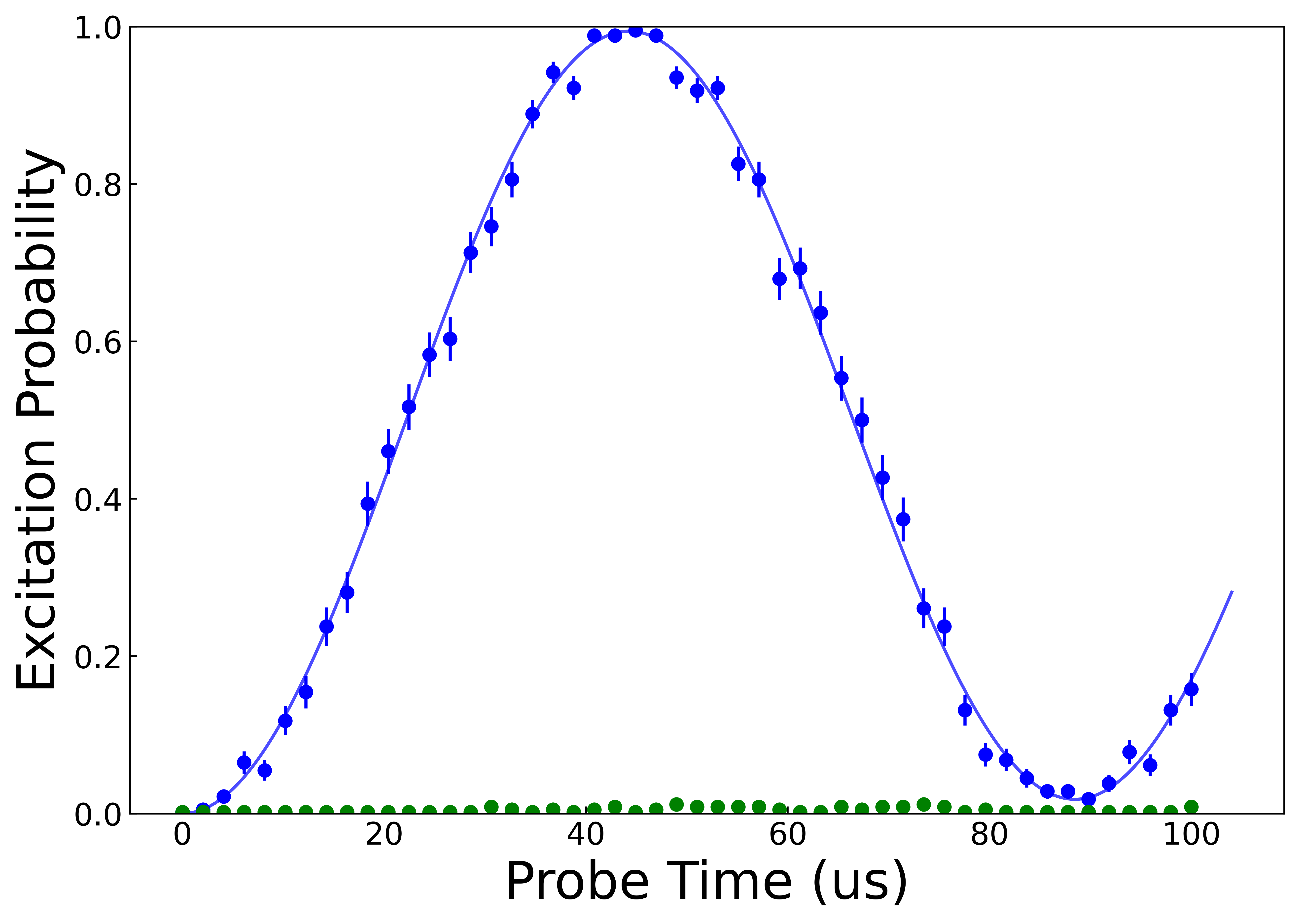}
    \label{fig:individual_addressing}}

    \caption{Micromotion-assisted individual addressing. (a) Schematics of the procedure. The cooling and memory ions are marked with "g" and "m" respectively. To address the memory ion, the two-ion chain is rotated around the cooling ion so that the Rabi frequency on the micromotion sideband is much higher for the memory ion. The dashed circle marks the position of the memory ion when micromotion is compensated. (b) Rabi flopping on the micromotion sideband with the rotated chain. The green and blue dots represent two-ion and one-ion excitation events respectively. The solid line represents fitting Rabi oscillation assuming a thermal state of motion.}
    \label{level_structure}
\end{figure*}

Micromotion-assisted individual addressing has been proposed \cite{Myatt1998} and demonstrated for a two-dimensional ion array \cite{Lysne2024}. In this section, we describe a realization in a two-ion chain. To address the memory ion, the two-ion chain is rotated about the cooling ion, perpendicularly to the trap surface, via controlled variation of the trap-electrode voltages (see Fig.~\ref{fig:twisting_chain}). When the memory ion is off-axis radially, it experiences a larger micromotion amplitude compared to the cooling ion. We add a 2~ms delay after the trap voltages are updated to account for low-pass filters present on the electrodes. We then drive the micromotion sideband of the S$_{1/2} \ket{F = 1, m_F = 0}$  to D$_{5/2} \ket{F = 2, m_F = 2}$ transition with a global beam. The $\pi$-time of the micromotion sideband for the memory ion is about 50~us (6~us for the carrier). Rabi flopping on the micromotion sideband using a global beam is shown in Fig.~\ref{fig:individual_addressing}. The fitted amplitude of oscillation is 99.9\%. Combined with state detection after the micromotion-based state preparation, the error for memory-ion preparation is negligible. After the memory ion is prepared in the $\ket{0}$ state, the ion chain is rotated back to the original position.

\subsection{Trap-induced qubit frequency fluctuations}
\label{app:RF_fluc}
\begin{figure}[!ht]
\includegraphics[width=3.4in]{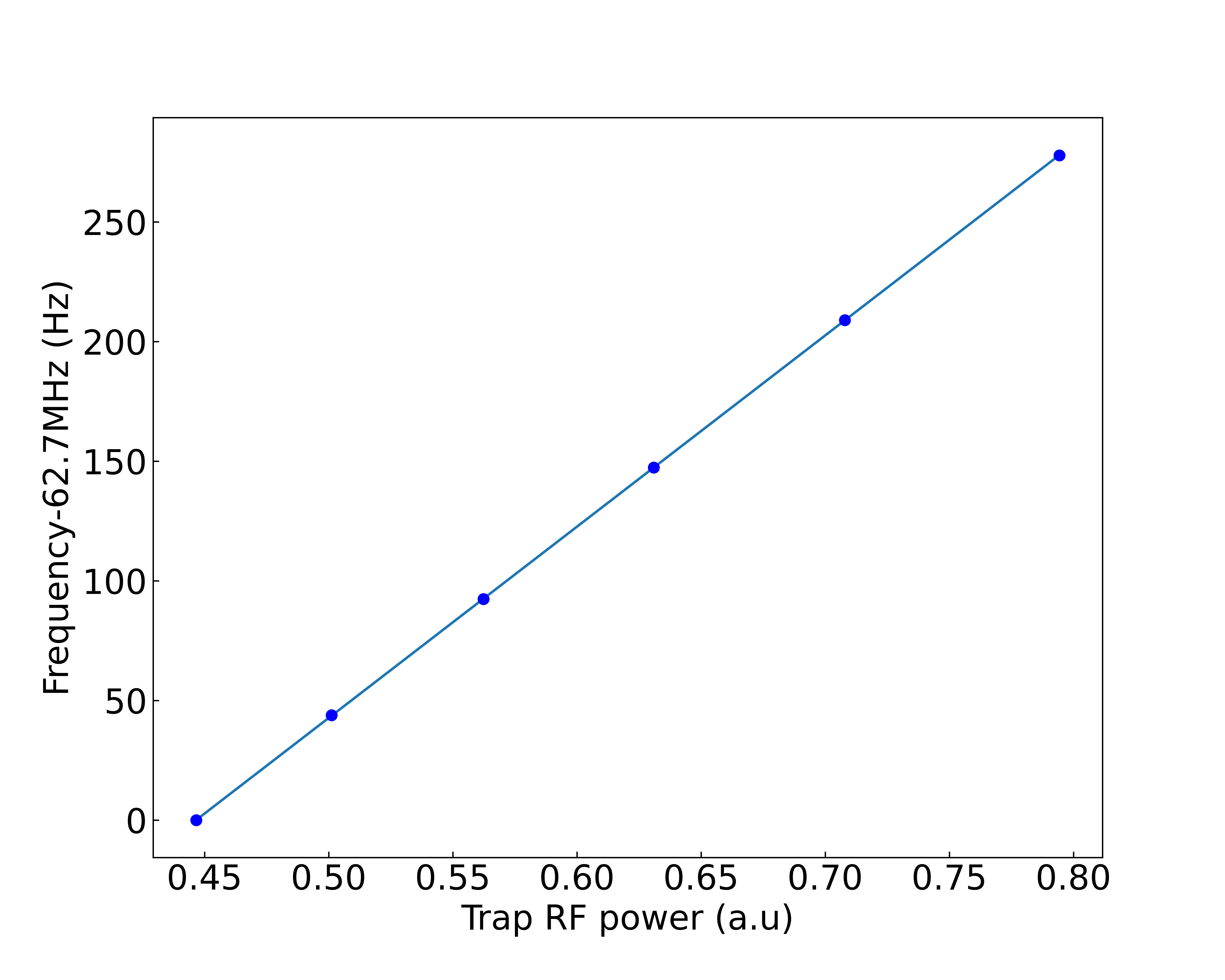}
\caption{The qubit frequency responds linearly to the output amplitude of the function generator that supplies the trapping field. The solid line is a linear fit to the data points. This shows that the AC Zeeman shift due to the oscillating trapping field is the dominant effect of the trap RF on the qubit. Slow fluctuations of the drive power could thus help explain the observed low-frequency noise. Error bars are smaller than the markers }
\label{fig:freq_vs_trap_amp}
\end{figure}

 The AC Zeeman shift due to the oscillating current in the RF electrodes scales quadratically with magnetic field, and thus linearly with the applied RF power. On the other hand, the quadrupole shift is caused by the gradient of the trapping field, and so scales with the square root of RF power. In a separate experiment, we measure the qubit transition frequency as a function of the trap-RF drive power. The result is shown in Fig.~\ref{fig:freq_vs_trap_amp}. The linear relation between the qubit frequency and RF power suggests that the AC Zeeman shift is likely the dominant source of trap-induced frequency shift. Thus, slow fluctuations of the trap RF amplitude could contribute to the measured low-frequency noise.
 
\subsection{Spontaneous scattering rate}
\label{app:scattering-rate}

Spontaneous Raman scattering is a source of errors affecting memory-qubit coherence, but since the predominant scattering channel is D$_{5/2} \rightarrow$ P$_{3/2} \rightarrow$ S$_{1/2}$, and since there is also decay to the D$_{3/2}$ level, many scattering errors are correctable via our leakage detection protocol~\cite{Kang2023}.  We estimate the scattering rate from the Doppler cooling beams on the D$_{5/2}$ qubit using the  Kramers-Heisenberg formula \cite{Ozeri_2005, Moore_2023}

\begin{equation}
    \Gamma_{i \rightarrow f} \approx \sum_{i, f} g^2 \gamma \left|\frac{a_{i \rightarrow f}^{(3/2)}}{\Delta} \right|^2.
\end{equation}

\noindent where $g = \frac{E_0\mu}{2\hbar}$, $\gamma$ is the decay rate of the P$_{3/2}$ manifold, $\mu$ is the relevant dipole matrix element $\langle $P$_{3/2}| \vec{d} \cdot \sigma_{l}|$D$_{5/2}\rangle$, $a_{i \rightarrow f}^{(3/2)} = \sum_q\sum_{k \in \rm P_{3/2}} \bra{f} \bold{d} \cdot \sigma_q \ket{k}\bra{k} \bold{d} \cdot \sigma_l \ket{i}/\mu^2$ is the effective transition amplitude for laser polarization $\sigma_l$ and scattered photon polarization $\sigma_q$, and $\Delta$ is the detuning from the D$_{5/2}$ - P$_{3/2}$ resonance. The measured optical power in the 493 nm (650 nm) beam is approximately 150~$\mu$W (20~$\mu$W), with a 20~$\mu$m (30~$\mu$m) beam diameter. 

From the reduced matrix element, we have that $\mu \approx 5$ a.u. \cite{Zhang2020} and $\sum_{i, f} |a_{i \rightarrow f}^{(3/2)}|^2 \sim 1$, which leads to a scattering rate of ${\sim} 2 \times 10^{-3}$~Hz, similar for both of these cooling beams.   For the undetectable scattering errors, those proceeding via D$_{5/2} \rightarrow$ P$_{3/2} \rightarrow$ D$_{5/2}$, however, the scattering rate can be bounded at a lower rate of ${\sim} 5 \times 10^{-4}$~Hz for each of these beams due to the branching ratio of the P$_{3/2} \rightarrow$ D$_{5/2}$ decay (0.23).  We therefore expect this effect to be a very small contribution to the total memory error in this demonstration.

\subsection{Dynamical-decoupling based noise filtering}
\label{app:cpmg}
Magnetic field noise arises from power fluctuations in the trapping RF and induces AC Zeeman shifts of the memory levels. Such fluctuations of the energy splitting are captured by including a dephasing term to the bare qubit splitting:
\begin{equation}
    H = \frac{\hbar}{2} \big(\omega_{0} + \beta(t)\big) \sigma_z.
    \label{tot-H}
\end{equation}

Frequency and phase noise in the driving RF pulses also lead to effective dephasing during the Ramsey sequence. This noise occurs due to local oscillator noise affecting the timing and frequency of driving RF pulses.
Furthermore, the near field effects of the trapping RF field include quadrupole terms which will off resonantly couple E2 transitions leading to higher order Stark shifts. The effect of such energy level shifts is again dephasing noise, though noise contributions from quadrupolar terms are a few orders of magnitude smaller than dipolar terms. 

Let us assume the noise $\beta(t)$ is a stationary process $\langle \beta(t' + t) \beta(t')\rangle = \langle \beta(t) \beta(0)\rangle$. Then, we may define the noise spectral density using the Wiener–Khinchin theorem as
 \begin{equation}
     \langle \beta(t' + t) \beta(t')\rangle = \frac{1}{2 \pi} \int_{-\infty}^{\infty}  S(\omega) e^{i \omega t}  d\omega. 
     \label{w-k-thm}
 \end{equation}

\noindent We will now connect the noise spectral density to the qubit coherence measured in the ion experiments to allow for noise spectrum determination, and subsequently for tailoring of the dynamical decoupling sequence for optimal noise rejection.

The advantage of CPMG pulses is that we can filter which components of noise spectra the ion experiences \cite{Biercuk_2011}. The CPMG sequence is comprised of $\pi$-pulses spaced by time $\tau$. If we assume the $\pi$-pulses are instantaneous, a good approximation for the long free-evolution times explored here, the time evolution to first-order perturbation approximation is:
\begin{equation}
\begin{split}
    U_{\rm CMPG} &= \prod_{j=0}^N e^{i \sigma_z \int_{t_j}^{t_{j+1}}\beta(t')\ dt'} e^{i \pi \sigma_y/2} \\
    & = e^{i \int_{-\infty}^{\infty} y(t, T) \beta(t) dt \sigma_z},
\end{split}
\end{equation}

\noindent where $t_j$ is the time of the $j$th $\pi$ pulse with $t_0 = 0, t_1 = \tau/2$ and $t_{N+1} = N \tau = T$. Here $y(t)$ alternates between $+1$ and $-1$ in successive free evolution periods, and the $\sigma_{i}$'s are the usual Pauli matrices. 

This periodic flipping of the Bloch vector is expressed using a modulation function $\Tilde{y}(\omega)$ in the frequency domain:
\begin{equation}
\begin{split}
     \Tilde{y}(\omega,T) &= \frac{1}{\sqrt{2 \pi}} \int_{0}^{\infty} y(t, T) e^{i \omega t} dt \\
     &= \frac{1}{\omega \sqrt{2 \pi}} \sum_{j=0}^N (-1)^j \left(e^{i \omega t_{j+1}} - e^{i \omega t_j}\right).
\end{split}
\end{equation}
The center of the filter function is at $\omega_c = 2\pi\times\frac{1}{2\tau}$, and the width is $1/T$. 

We calculate the coherence time of the memory qubit by performing a Ramsey protocol with or without a dynamical decoupling sequence during the Ramsey delay time:
\begin{equation}
    \langle \sigma_y \rangle = \frac{1}{2} 
    \left(1 + \frac {\langle e^{i \theta} + e^{-i \theta}\rangle }{2} \right) 
\end{equation}
 where $\theta(T) = \int_{-\infty}^{\infty} y(t, T) \beta(t) dt$ is the accumulated phase during total delay time $T$. The Ramsey contrast can be re-written using the cumulant expansion:
\begin{equation}
    \frac {\langle e^{i \theta} + e^{-i \theta}\rangle }{2}  = e^{-\chi(T)},
\end{equation}
 where the coherence $\chi(T) = \sum_k \frac{(-i)^k}{k!} C_k$ is the cumulant expansion of the noise with $C_k = \langle \theta^k(T) \rangle$ \cite{cumexp}.

We assume $\beta(t)$ is Gaussian (i.e. fully characterized by its auto-correlation), and so only the second cumulant survives $\chi(T) = \langle \theta^2(T) \rangle /2$.
Using Fubini's theorem~\cite{fubini1958}, we further simplify the integral for $\theta$ to obtain the coherence in the frequency domain:
\begin{equation}
    \begin{split}
    \chi(T) & = \frac{1}{2} \int_{-\infty}^{\infty}  dt' \int_{-\infty}^{\infty}  dt'' \langle \beta(t') \beta(t'') \rangle y(t') y(t'')\\
    & = \frac{1}{2} \int_{-\infty}^{\infty}  dt' \int_{-\infty}^{\infty}  dt'' \int_{-\infty}^{\infty} \frac{d\omega}{2 \pi} S(\omega) e^{i\omega(t'-t'')} y(t') y(t'')\\
    &= \int_{0}^{\infty}S(\omega)|\Tilde{y}(\omega)|^2 d\omega.
    \end{split}
\end{equation}
This gives a relation between coherence and noise spectral density via the filter function, allowing for translation of qubit measurements under a particular modulation scheme to a noise spectrum.

 The typical Ramsey protocol with modulation $y(t') = 1$ for $0<t'<T$ filters the noise as a sinc function with center about $0$ and width decreasing with delay time as $\sim 1/T$.
\begin{equation}
        \chi(T) =  \frac{1}{4\pi}\int_{-\infty}^{\infty} \frac{\sin^2(\omega \, T /2)}{(\omega/2)^2} S(\omega) d\omega.
        \label{chi-ramsey}
\end{equation}
Thus, the coherence for the typical Ramsey protocol is $\chi(T) = S(0) T / 2$ and is primarily due to the noise spectral component at $\omega = 0$.  We may however extract noise strength for other frequencies by using other modulating functions.

Considering only the main peak (first harmonic) of $\Tilde{y}(\omega)$ for more complex modulation sequences, we may approximate the noise spectrum from the measured Ramsey contrast. Using the definition of the $\delta$ function as $\delta(x) = \frac{1}{2\pi}\int_{-\infty}^{\infty} e^{i k x} dk$, we have that $\int_{-\infty}^{\infty} d\omega |\Tilde{y}(\omega)|^2 = \int_{-\infty}^{\infty} dt |y(t)|^2 = T$. Since the center of the filter function is at $\omega_c = 2\pi \times \frac{1}{2\tau}$, we may approximate the noise at $\omega_c$ from measured Ramsey contrast as
\begin{equation}
    \chi(T) = S(\omega_c)T/2;
    \label{flat_assump}
\end{equation}
there are higher harmonics to the filter function, but at lower amplitude, and we ignore them in this approximation.  We therefore can straightforwardly identify noise spectrum components to build up a spectrum by measuring ion coherence for experiments with different center frequencies and delay times.

\subsection{Maximum likelihood noise model}
\label{app:noise-modelling}

The CMPG sequence is used to measure the noise spectrum affecting the memory-qubit energy splitting \cite{Wang2017}. For these experiments, we keep the number of $\pi$-pulses fixed, varying the inter-pulse time $\tau$, and measure the coherence time.  We repeat this procedure with N = 2, 20, and 200 $\pi$-pulses. Using \eqref{flat_assump}, we can convert measured coherence times into noise spectral components \cite{DDcolornoise}. To get an approximate noise spectrum from the noise spectroscopy measurements (see Fig.~\ref{fig:modelled-noise}), we use a cost function to fit all the experimental data via maximum likelihood estimation.
We assume the noise exhibits power-law behavior and we include higher harmonics of the filter function up to $1$~MHz, parameterizing the spectrum as 
\begin{equation}
    S(\omega, \{\lambda_i\}) = \lambda_0 + \lambda_1 \omega + \lambda_2/ \omega + \lambda_3 / \omega^2,
\end{equation} 
where the $\lambda_{i}$'s represent the relative strength of the various power-law frequency-scaling terms in the fit function.  We generate coherences $\chi_{\rm num}$ from the fitted spectrum via
\begin{equation}
    \chi_{\rm num}(N, T, \{\lambda_i\}) = \int_{0}^{\infty}S(\omega,  \{\lambda_i\})|\Tilde{y}(\omega, N, T)|^2 d\omega.
\end{equation}
The cost function is given by weighted least squares error as
\begin{equation}     
    \mathcal{C}(\{\lambda_i\}) = \sum_{\{N, T\}} \frac{\left(\chi_{\rm exp}(N, T) - \chi_{\rm num}(N, T, \{\lambda_i\})\right)^2}{\sigma^2(N, T)};
\end{equation}
this compares the experimental coherence data $\chi_{\rm exp}(N, T)$ and respective variances $\sigma^2(N, T)$ from the N$=2, 20, 200$ measurements with varying delay time $T = N\tau$ to a given power law description $\chi_{\rm num}$. 
Optimizing this cost function leads to parameters $\lambda_0 = 0.0085, \lambda_1 = 0.0138, \lambda_2 = 0, \lambda_3 = 1.6631$. For these parameters, noise terms proportional to $1/\omega^2$ and $\omega$ dominate the spectrum. The transition point between these two noise scaling regimes is approximately $5$~Hz, and so we use a CPMG sequence centered about this minimum with $\tau = 0.1$~s to most effectively suppress the effect of the noise measured in the system. From the modeled noise spectrum, we calculate the expected qubit coherence time with erasure detection using~\ref{flat_assump} to be $115$~s, in reasonable agreement with the measured coherence time (cf. Fig.~\ref{fig:ramsey_135s}).

\end{document}